\documentclass[aps,prl,twocolumn,groupedaddress,showpacs]{revtex4}

\usepackage{graphicx}
\usepackage{dcolumn}
\usepackage{bm}

\begin{document}

\title{Magnetic-Field-Induced Localization of Quasiparticles in
Underdoped La$_{2-x}$Sr$_x$CuO$_4$ Single Crystals}

\author{X. F. Sun}
\email[]{ko-xfsun@criepi.denken.or.jp}
\author{Seiki Komiya}
\author{J. Takeya}
\author{Yoichi Ando}
\email[]{ando@criepi.denken.or.jp}
\affiliation{Central Research
Institute of Electric Power Industry, Komae, Tokyo 201-8511,
Japan.}

\date{\today}

\begin{abstract}

Magnetic-field-induced ordering of electrons around vortices is a
striking phenomenon recently found in high-$T_c$ cuprates. To
identify its consequence in the quasiparticle dynamics, the
magnetic-field ($H$) dependence of the low-temperature thermal
conductivity $\kappa$ of La$_{2-x}$Sr$_x$CuO$_4$ crystals is
studied for a wide doping range. It is found that the behavior of
$\kappa(H)$ in the sub-Kelvin region changes drastically across
optimum doping, and the data for underdoped samples are
indicative of unusual magnetic-field-induced localization of
quasiparticles; this localization phenomenon is probably
responsible for the unusual ``insulating normal state" under high
magnetic fields.

\end{abstract}


\maketitle

Competing order is an important emerging concept in high-$T_c$
superconducting cuprates \cite{Sachdev}. Recently, Hoffman {\it et al.}
\cite{Hoffman} reported scanning tunneling microscopy (STM)
studies of Bi$_2$Sr$_2$CaCu$_2$O$_{8+\delta}$ (Bi2212) single
crystals in magnetic fields, which revealed a ``checkerboard-like"
order of quasiparticle (QP) states with four-unit-cell
periodicity surrounding vortex cores. This result documented a
magnetic-field-induced charge density wave (CDW), which is in
good correspondence with the magnetic-field-induced spin density
wave (SDW) found by neutron scattering in La$_{2-x}$Sr$_x$CuO$_4$
(LSCO) \cite{Lake1,Lake2} and La$_2$CuO$_{4+\delta}$
\cite{Khaykovich}, and these phenomena are likely to be results
of competing antiferromagnetic and superconducting orders in
those materials. In fact, several theoretical models
\cite{Demler,Franz,Zhu,Chen,Kivelson} have been proposed to
describe the coexistence of $d$-wave superconductivity and
spin/charge order in the vortex state as a consequence of
competing orders. However, it is hardly known how such
competition and magnetic-field-induced order affect the QP
dynamics, the details of which would allow us to understand the
nature of the novel magnetic-field induced states.

In this Letter, we report that the magnetic-field dependence of
thermal conductivity, $\kappa(H)$, of LSCO at low temperature
demonstrates that the magnetic-field-induced order leads to
unusual localization of QPs; moreover, we found that there is a
distinct change in the behavior of $\kappa(H)$ near optimum
doping and the QP localization occurs only in the underdoped
samples. This magnetic-field-induced localization of QPs in the
underdoped regime is clearly in correspondence with the
well-known metal-to-insulator crossover, found in the
low-temperature normal state under 60-T magnetic field, that
occurs at optimum doping \cite{Boebinger}. Moreover, this result
suggests that the unusual ``insulating normal state" of the
cuprates under 60 T, that is characterized by a log(1/$T$)
divergence in resistivity \cite{Ando1}, is caused by the
magnetic-field-induced order.

In high-$T_c$ cuprates, it is established that the
superconducting gap $\Delta$ has essentially the $d_{x^2-y^2}$
symmetry, which has four nodes (where the gap magnitude vanishes)
along the diagonals of the square Brillouin zone. Because of
these nodes, QPs are easily created both by thermal fluctuations
and by the impurity scattering, and the thermal conductivity
$\kappa$ is a useful bulk probe of the QP excitations and their
transport behavior \cite{Durst,Kubert}. In the mixed state, where
the magnetic field enters the superconductor in the form of
quantized vortices, it is known that the magnetic field induces
extended zero-energy QP states near the gap nodes and their
population increases as $H^{1/2}$ \cite{Durst,Kubert,Moler}; this
so-called ``Volovik effect" is due to the Doppler shift of the QP
energy spectrum in the presence of a supercurrent flowing around
each vortex \cite{Volovik}. As a result, the QP heat transport is
enhanced in magnetic fields at low temperatures (usually in the
sub-Kelvin region) \cite{Chiao,Aubin,Ando2,Proust}, while at
higher temperatures the QP heat transport is suppressed in
magnetic fields because of the dominance of vortex scattering of
QPs \cite{Krishana,Ong,Ando3}. Interestingly, at intermediate
temperatures a ``plateau" shows up in the $\kappa(H)$ profile of
Bi2212 \cite{Krishana}, which, after a long debate (see Ref.
\cite{Ando3} and references therein), can be understood to be a
result of the competition between the increase in the QP
population and the decrease in their mean free path with $H$
\cite{Ando2}. Till now, the magnetic-field-induced enhancement of
QP transport was studied in optimally-doped
YBa$_2$Cu$_3$O$_{7-\delta}$ (YBCO) \cite{Chiao} and Bi2212
\cite{Aubin,Ando2} and in overdoped Tl$_2$Ba$_2$CuO$_{6+\delta}$
\cite{Proust}, but its doping dependence has been scarcely known.

Here we choose to study the low-temperature heat transport of
LSCO, in which the hole doping can be well controlled over a wide
range. High-quality La$_{2-x}$Sr$_x$CuO$_4$ single crystals ($x$
= 0.08, 0.10, 0.14, 0.17 and 0.22) are grown by the
traveling-solvent floating-zone technique \cite{Ando4}. The
crystals are cut into rectangular platelets with typical size of
$1.5 \times 0.5 \times 0.1$ mm$^3$, where the $c$ axis is
perpendicular to the platelets within an accuracy of 1$^{\circ}$,
determined by the X-ray Laue analysis. The samples at $x$ =
0.08--0.17 are annealed at 800--880 $^{\circ}$C in air for 60
hours, followed by rapid quenching to room temperature, to remove
oxygen defects, while the sample at $x$ = 0.22 is annealed in
oxygen and quenched to 77 K. Magnetic susceptibility measurements
show that the superconducting transition temperature $T_c$ is 22,
28.5, 35.5, 36.5 and 28 K for $x$ = 0.08, 0.10, 0.14, 0.17 and
0.22, respectively. The zero-field heat transport in these
samples at milli-Kelvin temperatures has already been reported in
a previous paper \cite{Takeya}. In this work, both the
temperature and the magnetic-field dependences of $\kappa$ are
measured from 0.3 to 7 K in a $^3$He refrigerator by using the
conventional steady-state ``one heater, two thermometer"
technique \cite{Ando2}. The magnetic field up to 16 T is applied
along the $c$ axis of the crystals while the heat current flows
in the $ab$ plane.  To avoid complications that are associated
with the vortex-pinning-related hysteresis \cite{Aubin}, the
$\kappa(H)$ data are taken in the field-cooled procedure
\cite{Ando2,Ando3}.

\begin{figure}
\includegraphics[clip,width=6.5cm]{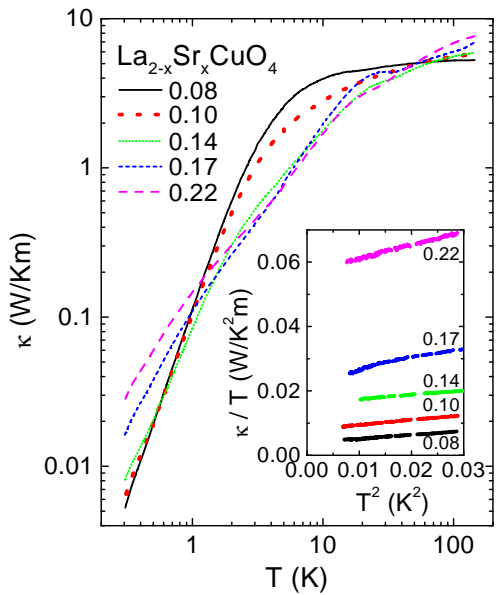}
\caption{Temperature dependence of the in-plane thermal
conductivity $\kappa$ in zero magnetic field for
La$_{2-x}$Sr$_x$CuO$_4$ single crystals with various hole doping.
The inset shows the data in milli-Kelvin region \cite{Takeya},
plotted in $\kappa/T$ vs. $T^2$, which clearly shows the
enhancement of QP heat transport with $x$.}
\end{figure}

Figure 1 shows the zero-field $\kappa(T)$ of LSCO crystals in a
wide temperature range, where the additional data from 5 to 150 K
are taken using a Chromel-Constantan thermocouple in a $^4$He
cryostat, and the very low temperature $\kappa(T)$ data
\cite{Takeya} are also shown in the inset. The measured thermal
conductivity is a sum of the electron (or QP) term $\kappa_e$ and
the phonon term $\kappa_{ph}$, whose temperature dependences are
proportional to $T$ and $T^3$ in the low-$T$ limit,
respectively.  Thus, in the $\kappa$/$T$ vs $T^2$ plot (Fig. 1
inset), it is clear that the $\kappa_e$ component increases with
$x$. In the main panel of Fig. 1, the superconducting transition
can barely be recognized by a weak hump in $\kappa$, which is
easiest to see in the optimally-doped sample ($x$ = 0.17).

\begin{figure}
\includegraphics[clip,width=8.5cm]{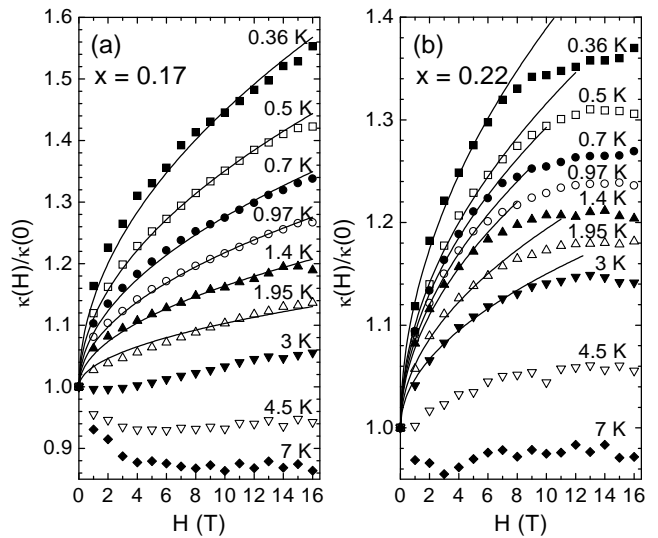}
\caption{Magnetic-field dependences of the thermal conductivity
$\kappa$ for the optimum (a) and overdoped (b) LSCO crystals ($x$
= 0.17 and 0.22, respectively). Solid lines in each panel
show fits of the low-$T$ data to the $H^{1/2}$ dependence, which
is expected when the magnetic-field enhancement in $\kappa$ is
mainly due to the increase in the QP population by the Volovik
effect.}
\end{figure}

Figure 2 shows the magnetic-field dependences of $\kappa$ for
optimally-doped ($x$ = 0.17) and overdoped ($x$ = 0.22) LSCO.
The optimally-doped sample shows a $\kappa(H)$ behavior very
similar to that in Bi2212 \cite{Aubin,Ando2} and YBCO
\cite{Chiao,Ong}; namely, $\kappa$ is enhanced with $H$ at very
low temperatures, but is suppressed with $H$ at higher
temperatures. Not surprisingly, a clear ``plateau" shows up in
the $\kappa(H)$ profile at 7 K. It is useful to note that the
low-$T$ $\kappa(H)$ data of the optimally-doped sample can
roughly be described by the formula
$\kappa(H)/\kappa(0)=1+aH^{1/2}$ ($a$ is a temperature-dependent
parameter), as is shown in Fig. 2(a). This $H$-dependence is
expected \cite{Aubin} if the vortex scattering is negligibly
small and the rise in $\kappa$ is attributed simply to the
increase in QP population due to the Volovik effect. (Note that
the phononic contribution to $\kappa$ is believed to be
independent of $H$ in the cuprates
\cite{Chiao,Aubin,Ando2,Proust,Krishana,Ong,Ando3}.) The
reasonable fits in Fig. 2(a) demonstrate that the vortex
scattering becomes relatively insignificant at low temperatures
at optimum doping. In overdoped LSCO ($x$ = 0.22), the general
trend of $\kappa(H)$ is similar, although there is some
difference at high fields; as is shown in Fig. 2(b), the behavior
of $\kappa(H)$ follows $H^{1/2}$ law at low fields but noticeably
deviates from $H^{1/2}$ at high fields, presenting a
``plateau-like" feature. This flattening of $\kappa(H)$ probably
indicates an increasing importance of QP scattering at high
fields. One possible cause of this feature is the effect of
inelastic electron-electron scattering, which may remain
important at very low temperatures in overdoped samples
\cite{Mackenzie}. In any case, the magnetic-field dependence of
the QP heat transport in optimum and overdoped LSCO is
qualitatively similar to that in the optimally-doped Bi2212 and
YBCO.

\begin{figure*}
\includegraphics[clip,width=16cm]{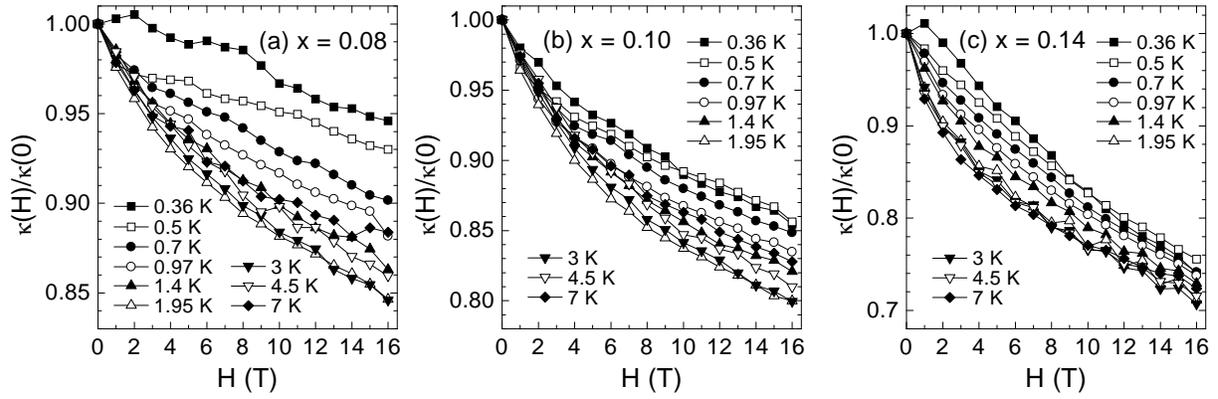}
\caption{Magnetic-field dependences of $\kappa$ for the underdoped LSCO
crystals ($x$ = 0.08, 0.10, and 0.14) at various temperatures down to
0.36 K. Note that the magnetic-field enhancement of $\kappa$ is gone in
those underdoped samples.}
\end{figure*}

The most surprising result of the present work is the behavior of
$\kappa(H)$ in underdoped LSCO ($x$ = 0.08, 0.10 and 0.14).  As
shown in Fig. 3, $\kappa$ always decreases quickly with $H$ at
all temperatures down to 0.36 K in all the underdoped crystals
studied, which is vastly different from the $\kappa(H)$ behavior
of optimum and overdoped crystals.  To understand the implication
of this result, it should first be recognized that the impurity
scattering cannot be the source of such a strong magnetic-field
suppression of QP heat transport, because it has been
demonstrated \cite{Ando3} that impurities just strongly {\it
diminishes} the $H$ dependence of $\kappa$.  Therefore, the
effect of magnetic field on the QP heat transport must have been
changed fundamentally in underdoped samples. When the data in
Fig. 3 are compared to those in Fig. 2, it is obvious that both
the magnetic-field enhancement of $\kappa$ at low $T$ and the
``plateau" feature at high $T$ are absent in underdoped samples,
which points to an increased role of vortex scattering and a
diminished role of magnetic-field-induced QPs. Given that the
specific heat measurements have already demonstrated
\cite{SJChen} that the QP density-of-states are always enhanced
with $H$ (i.e., QPs are certainly created by magnetic fields) in
underdoped LSCO, it is most reasonable to conclude the following:
{\it (i) the magnetic-field-induced QPs are localized and
contribute little to the heat transport in underdoped LSCO, and
(ii) vortices strongly scatter QPs in underdoped samples even at
low $T$, while they do not effectively scatter QPs in optimum and
overdoped samples at low $T$.}

These results are most naturally understood in the light of the
recently-found magnetic-field-induced CDW/SDW in the cuprates
\cite{Hoffman,Lake1,Lake2,Khaykovich}. For LSCO, neutron
scattering experiments have shown \cite{Lake2,Khaykovich} that the
magnetic-field-induced SDW is dynamical at optimum doping, while
it becomes a static, zero-energy object in underdoped samples.
Thus, the magnetic-field-induced order is expected to be relevant
to the low-temperature properties only in underdoped LSCO.
Moreover, the STM study of Bi-2212 found checkerboard-like CDW
around vortices \cite{Hoffman}, which is naturally expected to
cause enhanced QP scattering. Therefore, it is likely that in
underdoped LSCO the magnetic-field-induced order is responsible
for both the QP localization and the enhanced QP scattering off
the vortices. It is thus reasonable to conclude that the peculiar
$\kappa(H)$ behavior of underdoped LSCO points to novel
magnetic-field-induced localization of QPs that is caused by the
magnetic-field-induced order around vortices.

\begin{figure}
\includegraphics[clip,width=6.0cm]{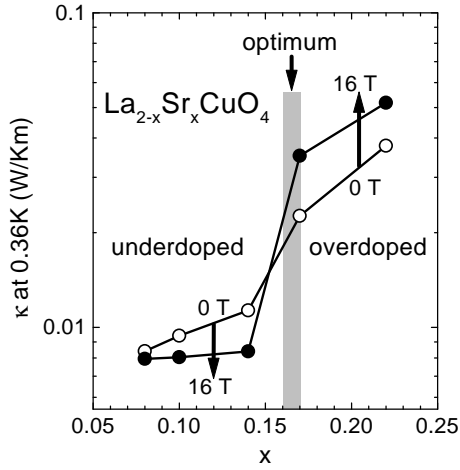}
\caption{Hole-doping dependence of the thermal conductivity
$\kappa$ at 0.36 K in 0 T (open circles) and in 16 T (solid
circles). $\kappa$ in 16 T shows a pronounced jump at optimum
doping (marked by a gray band), which demonstrates that the role
of magnetic field changes drastically at optimum doping. The
suppression of $\kappa$ with $H$ in the underdoped regime is a
result of the magnetic-field-induced localization of QPs.}
\end{figure}

Figure 4 shows the hole-doping dependence of $\kappa$ at 0.36 K
in 0 and 16 T, which summarizes our main finding.  In 0 T,
$\kappa$ shows a relatively smooth change with $x$, but $\kappa$
in 16 T  shows a pronounced jump at optimum doping. As one can
see in Fig. 4, this jump is caused by the fact that $\kappa$ is
{\it suppressed} with $H$ in the underdoped regime, while it is
{\it enhanced} with $H$ in the optimum and overdoped regime; we
discussed that the former is caused by the magnetic-field-induced
localization of QPs that is likely associated with the competing
order, while the latter reflects extended QPs created by magnetic
fields.  Therefore, Fig. 4 depicts the fact that the role of
magnetic field changes drastically at optimum doping.

An intriguing implication of the present result is that the
magnetic-field-induced QP localization may also be responsible
for the ``insulating" normal-state resistivity under 60 T
\cite{Boebinger,Ando1}. The fact that the ``insulating normal
state" under 60 T is observed only in underdoped samples
\cite{Boebinger} is strongly suggestive of the correspondence. In
fact, it is easy to imagine that the charge-ordered region around
vortices completely overlap in 60-T field and all the electrons
are incorporated into the magnetic-field-induced ordered state.
Thus, it is tempting to assert that the log(1/$T$) behavior of
the resistivity under 60 T \cite{Ando1} is a property of the
state where the magnetic-field-induced order proliferates.
In this regard, it is worth noting that a recent Nernst effect
measurement found a sizable Nernst signal in the ``insulating
normal state" in high magnetic fields \cite{Capan}.
In fact, the behavior of the Nernst signal below $T_c$ is not much
different between underdoped and overdoped LSCO, which probably reflects
the fact that the observed Nernst signal primarily comes from the vortex
motion and is insensitive to the QP contribution \cite{Capan,Wang}.
Incidentally, the Nernst effect has been utilized to measure the upper
critical field $H_{c2}$ of cuprates, where $H_{c2}$ was found to show a
rather smooth change across optimum doping \cite{Wang}; this fact rules
out the possibility that the observed change in $\kappa(H)$ between $x$
= 0.14 and 0.17 might be caused by a sudden change in $H_{c2}$, which
would cause the reduced field $H/H_{c2}$ to vary drastically.

It is useful to note that our previous results in zero magnetic
field have shown \cite{Takeya} that the residual electronic
thermal conductivity divided by $T$, $\kappa_{\rm res}/T$, is
much larger than what the normal-state resistivity under 60 T
would suggest in the underdoped LSCO, which may indicate a strong
violation of the Wiedemann-Franz law. The present new data show
that in underdoped LSCO the QP transport is strongly suppressed
with $H$ even at low $T$, and thus it is expected that the
$\kappa_{\rm res}/T$ value in high magnetic field may smoothly
match the value estimated from the normal-state resistivity under
60 T. This is an indication that the magnetic-field-induced order
around vortices ultimately changes the global properties of
underdoped LSCO in high magnetic field. Therefore, the thermal
transport properties give evidence that the magnetic field can
cause a drastic change in the low-energy physics of underdoped
cuprates because of the magnetic-field-induced localization of
QPs, which is likely related to the inherent competition between
antiferromagnetic and superconducting orders.

In summary, we measure the magnetic-field dependences of the
thermal conductivity of LSCO single crystals for a wide doping
range. It is found that the $\kappa(H)$ behavior at low-$T$ gives
evidence for novel magnetic-field-induced localization of QPs in the
underdoped regime. We discuss that this phenomenon is associated
with the magnetic-field-induced spin/charge order, and is probably
responsible for the unusual ``insulating normal state" under high
magnetic fields.

\begin{acknowledgments}
We thank S. A. Kivelson, A. N. Lavrov, and P. A. Lee for helpful
discussions.
\end{acknowledgments}


\begin{thebibliography}{}

\bibitem{Sachdev}
S. Sachdev and S. C. Zhang, Science {\bf 295}, 452 (2002).

\bibitem{Hoffman}
J. E. Hoffman {\it et al.}, Science {\bf 295}, 466 (2002).

\bibitem{Lake1}
B. Lake {\it et al.}, Science {\bf 291}, 1759 (2001).

\bibitem{Lake2}
B. Lake {\it et al.}, Nature (London) {\bf 415}, 299 (2002).

\bibitem{Khaykovich}
B. Khaykovich {\it et al.}, Phys. Rev. B {\bf 66}, 014528 (2002).

\bibitem{Demler}
E. Demler, S. Sachdev, and Y. Zhang, Phys. Rev. Lett. {\bf 87},
067202 (2001).

\bibitem{Franz}
M. Franz, D. E. Sheehy, and Z. Te\v{s}anovi\'{c}, Phys. Rev.
Lett. {\bf 88}, 257005 (2002).

\bibitem{Zhu}
J.-X. Zhu, I. Martin, and A. R. Bishop, Phys. Rev. Lett. {\bf
89}, 067003 (2002).

\bibitem{Chen}
H.-D. Chen, J.-P. Hu, S. Capponi, E. Arrigoni, and S.-C. Zhang,
Phys. Rev. Lett. {\bf 89}, 137004 (2002).

\bibitem{Kivelson}
S. A. Kivelson, D. H. Lee, E. Fradkin, and V. Oganesyan, Phys.
Rev. B {\bf 66}, 144516 (2002).

\bibitem{Boebinger}
G. S. Boebinger {\it et al.}, Phys. Rev. Lett. {\bf 77}, 5417
(1996).

\bibitem{Ando1}
Y. Ando, G. S. Boebinger, A. Passner, T. Kimura, and K. Kishio,
Phys. Rev. Lett. {\bf 75}, 4662 (1995).

\bibitem{Durst}
A. C. Durst and P. A. Lee, Phys. Rev. B {\bf 62}, 1270 (2000).

\bibitem{Kubert}
C. K\"{u}bert and P. J. Hirschfeld, Phys. Rev. Lett. {\bf 80},
4963 (1998).

\bibitem{Moler}
K. A. Moler {\it et al.}, Phys. Rev. Lett. {\bf 73}, 2744 (1994).

\bibitem{Volovik}
G. E. Volovik, JETP Lett. {\bf 58}, 469 (1993).

\bibitem{Chiao}
M. Chiao {\it et al.}, Phys. Rev. Lett. {\bf 82}, 2943 (1999).

\bibitem{Aubin}
H. Aubin, K. Behnia, S. Ooi, and T. Tamegai, Phys. Rev. Lett.
{\bf 82}, 624 (1999).

\bibitem{Ando2}
Y. Ando, J. Takeya, Y. Abe, X. F. Sun, and A. N. Lavrov, Phys.
Rev. Lett. {\bf 88}, 147004 (2002).

\bibitem{Proust}
C. Proust, E. Boaknin, R. W. Hill, L. Taillefer, and A. P.
Mackenzie, Phys. Rev. Lett. {\bf 89}, 147003 (2002).

\bibitem{Krishana}
K. Krishana, N. P. Ong, Q. Li, G. D. Gu, and N. Koshizuka,
Science {\bf 277}, 83 (1997).

\bibitem{Ong}
N. P. Ong {\it et al.}, in {\it Physics and Chemistry of
Transition Metal Oxides}, edited by H. Fukuyama and
N. Nagaosa (Springer-Verlag, Berlin, 1999), p. 202.

\bibitem{Ando3}
Y. Ando, J. Takeya, Y. Abe, K. Nakamura, and A. Kapitulnik, Phys.
Rev. B {\bf 62}, 626 (2000).

\bibitem{Ando4}
Y. Ando, A. N. Lavrov, S. Komiya, K. Segawa, and X. F. Sun, Phys.
Rev. Lett. {\bf 87}, 017001 (2001).

\bibitem{Takeya}
J. Takeya, Y. Ando, S. Komiya, and X. F. Sun, Phys. Rev. Lett.
{\bf 88}, 077001 (2002).

\bibitem{Mackenzie}
A. P. Mackenzie, S. R. Julian, D. C. Sinclair, and C. T. Lin,
Phys. Rev. B {\bf 53}, 5848 (1996).

\bibitem{SJChen}
S. J. Chen {\it et al.}, Phys. Rev. B {\bf 58}, R14753 (1998).

\bibitem{Capan}
C. Capan {\it et al.}, Phys. Rev. Lett. {\bf 88}, 056601 (2002).

\bibitem{Wang}
Y. Wang {\it et al.}, Science {\bf 299}, 86 (2003).

\end{thebibliography}
\end{document}